\documentclass[useAMS,usenatbib]{mn2e}
\usepackage{epsfig,natbib}
\usepackage{amssymb}
\usepackage{graphicx}
\usepackage{placeins}
\usepackage{times}
\usepackage{epsfig}
\newcommand\ion[2]{#1{\sc #2}}
\def\Zsun{\mbox{$Z_\odot$}}

\def\Vout{\mbox{$V_{\mathrm{out}}$}}
\newcommand{\lya}{Ly$\alpha$}
\def\kms{\mbox{km~s$^{-1}$}}

\newcommand{\NHI}   {\ensuremath{N_{\textrm{{\scriptsize H}{\tiny \hspace{.1mm}I}}}}}
\newcommand{\ave}[1]{\ensuremath{\langle #1 \rangle}}
\newcommand{\Ncl}   {\ensuremath{N_{\mathrm{cl}}}}

\title[Stellar population of a $z=2.35$ DLA galaxy]
{Comprehensive Study of a \boldmath{$z=2.35$} DLA Galaxy:\\
Mass, Metallicity, Age, Morphology and SFR from HST and VLT\thanks{
Based on observations collected at the European Organisation
for Astronomical Research in the Southern Hemisphere, Chile, under program
087.A-0085(A). Based on observations made
with the NASA/ESA Hubble Space Telescope, obtained at
the Space Telescope Science Institute, which is operated by the Association of
Universities for Research in Astronomy, Inc., under NASA contract NAS 5-26555.
These observations are associated with program 12553.}
}

\author[J.-K. Krogager et al.]{Jens-Kristian Krogager,$^{1,2}$\thanks{E-mail:
krogager@dark-cosmology.dk}
Johan P. U. Fynbo$^{1}$,
C\'edric Ledoux$^{2}$,
Lise Christensen$^{1}$,\newauthor
Anna Gallazzi$^{3,1}$,
Peter Laursen$^{1}$,
Palle M\o ller$^{4}$,
Pasquier Noterdaeme$^{5}$,\newauthor
C\'eline P\'eroux$^{6}$,
Max Pettini$^{7}$,
Marianne Vestergaard$^{1,8}$
\\
$^{1}$Dark Cosmology Centre, Niels Bohr Institute, Copenhagen University,
Juliane Maries Vej 30, 2100 Copenhagen \O, Denmark\\
$^{2}$European Southern Observatory, Alonso de C\'ordova 3107, Vitacura, Casilla 19001, Santiago 19,
Chile\\ 
$^{3}$INAF -- Osservatorio Astrofisico di Arcetri,
Largo Enrico Fermi 5, 50125 Firenze, Italy\\
$^{4}$European Southern Observatory, Karl-Schwarzschild-Strasse 2,
D-85748 Garching bei M\"unchen, Germany\\ 
$^{5}$UPMC-CNRS, UMR7095, Institut
d'Astrophysique de Paris, F-75014 Paris, France\\
$^{6}$Aix Marseille Universit\'e, CNRS, LAM
(Laboratoire d'Astrophysique de Marseille) UMR 7326, 13388, Marseille, France \\
$^{7}$Institute of Astronomy, Kavli Institute for Cosmology, Madingley
Road Cambridge, CB3 0HA\\ 
$^{8}$Steward Observatory, University of Arizona, 933 N Cherry Avenue, Tucson, AZ 85721, USA
}

\begin{document}

\date{Accepted 30 May 2013. Received ; in original form }

\pagerange{\pageref{firstpage}--\pageref{lastpage}} \pubyear{2002}

\maketitle

\label{firstpage}

\begin{abstract}
We present a detailed study of the emission from a $z=2.35$ galaxy that causes
damped Lyman-$\alpha$ absorption in the spectrum of the background QSO,
SDSS J 2222$-$0946. We present the results of extensive analyses of the
stellar continuum covering the rest frame optical-UV regime based on
broad-band {\it HST} imaging, and of spectroscopy from VLT/X-Shooter of the strong emission
lines: Ly$\alpha$, [\ion{O}{ii}], [\ion{O}{iii}], [\ion{N}{ii}], H$\alpha$ and H$\beta$.
We compare the metallicity from the
absorption lines in the QSO spectrum with the oxygen abundance inferred from
the strong-line methods (R$_{23}$ and N2). 
The two emission-line methods yield consistent results: [O/H] = $-0.30\pm0.13$.
Based on the absorption lines in the QSO spectrum a metallicity of
$-0.49\pm0.05$ is inferred at an impact parameter of 6.3 kpc from the centre of the galaxy
with a column density of hydrogen of $\log(\NHI / {\rm cm}^{-2})=20.65\pm0.05$. 
The star formation rates of the galaxy from the UV continuum and
H$\alpha$ line can be reconciled assuming an amount of reddening of
$E(B-V) = 0.06\pm0.01$, giving an inferred SFR of
$13\pm1~\mathrm{M}_{\odot}$~yr$^{-1}$ (Chabrier IMF). From the {\it HST}
imaging, the galaxy associated with the absorption is found to be a compact ($r_e$=1.12~kpc) object
with a disc-like, elongated (axis ratio 0.17) structure indicating that 
the galaxy is seen close to edge-on. Moreover, the absorbing gas is located
almost perpendicularly above the disc of the galaxy suggesting that the gas
causing the absorption is not co-rotating with the disc. We investigate the
stellar and dynamical masses from SED-fitting and emission-line widths, respectively, and find
consistent results of $2 \times10^9$~M$_{\odot}$. We suggest that the galaxy is a
young {\it proto}-disc with evidence for a galactic outflow of enriched gas.
This galaxy hints at how star-forming galaxies
may be linked to the elusive population of damped Ly$\alpha$ absorbers.

\end{abstract}


\section{Introduction}

Mapping the structure, properties and chemical enrichment of galaxies over cosmic
history is a major goal of contemporary astrophysics. In the local Universe, important
advances can be made by studying the age-metallicity relation of stars in the Solar
neighbourhood \citep[e.g.,][]{Holmberg2007,Caffau2011} or in local group dwarf
galaxies \citep{Frebel2007,Frebel2010}. At redshifts higher than $z\approx2$,
a powerful method for studying chemical evolution is spectroscopy
of Damped Ly$\alpha$ Absorbers (DLAs) detected either towards background QSOs
\citep[see][and references therein]{Wolfe05} or against Gamma-ray Burst (GRB) afterglow light
\citep{Fynbo06,Savaglio06,Prochaska07}. In those studies, abundances are determined from
the gas phase as probed by the H{\sc \,i} and metal absorption lines detected
against the light of the background QSOs and GRB afterglows, respectively. 

Mainly at lower redshifts, but recently also at $z\gtrsim2$,
H{\sc \,ii}-region abundances are determined using the relative strengths
of strong emission lines \citep[e.g.,][and references therein]{Kewley08,
Shapley11}.
Currently, there have only been very few cases where both methods have been
applied to the same object. The first example is the case of
the DLA towards SBS 1543+593 \citep{Bowen2005} where the two methods yielded
consistent results. Since then a handful of other cases have been studied
\citep[see the compilation in][]{Peroux12}. As seen in Fig.~8 of \citet{Peroux12},
the absorption-line measurements generally probe regions at larger galactocentric
distances than the emission-line based measurements and
on average indicate lower metallicities than the results based on emission-lines,
possibly reflecting the early setup of metallicity gradients \citep{oRourke2011}.
However, it is important to stress that different recipes to derive oxygen abundances
using strong emission-line fluxes reveal very inconsistent results \citep{Kewley08}.
It is therefore of interest to expand the sample of sources where strong-line
based abundances can be independently tested using other methods
\citep[see also][]{Pettini2006, Kudritzki12}. Also, the use of different elements, e.g.,
Fe, Zn or Si, to infer absorption metallicities may introduce systematic
offsets when comparing galaxies within a heterogeneous sample.

Combining the two complementary methods of studying the metal enrichment in
galaxies provides important hints to understanding how galaxies turn their gas
into stars, as the absorption lines directly probe the cold gas, and
the properties of the star forming region can be probed directly from the 
emission lines, and sometimes also from the continuum. However, linking the
absorption characteristics of DLAs to the emission characteristics of the
galaxies causing the absorption has been a great challenge at high redshift, due to the faint
nature of DLA galaxies and due to their proximity to a very bright QSO. So far,
only a few DLA systems with confirmed emission counterparts have been
established at redshifts around and higher than two \citep{Krogager2012}.

Here we present an analysis of the DLA at $z=2.35$ with $\log(\NHI / {\rm cm}^{-2})=20.65\pm0.05$
seen in the spectrum of the
$z=2.93$ QSO 2222$-$0946. The galaxy counterpart of the absorber was detected by
\citet{Fynbo10} and here we present new, deeper spectroscopic data allowing
us to measure the metallicity directly from nebular $[$O{\sc \,ii}$]$,
$[$O{\sc \,iii}$]$, and Balmer emission lines. Furthermore, we combine the
spectroscopic data with imaging from Wide Field Camera 3 (WFC3) onboard the
{\it Hubble Space Telescope} ({\it HST}). The high resolution images from {\it HST}
makes it possible to detect the continuum emission from the galaxy directly, allowing us to
characterize the properties and structure of the absorbing galaxy. This enables
us to start bridging the gap between the population of DLA galaxies and
star-forming galaxies.

The paper is organized as follows. In Sect.~2, we
present the spectroscopic data from X-Shooter and the imaging data from
{\it HST}/WFC3. In Sect.~3, we derive the oxygen abundance from nebular emission
lines, derive the gas phase metal abundances, characterize the morphology from
the imaging data, and finally we obtain fluxes of the galaxy used for spectral
energy distribution (SED) fitting. In Sect.~4, we compare the absorption and emission
properties, and discuss and outline the implications of our work. Throughout
this paper, we assume a standard $\Lambda$CDM cosmology with $H_0=71\,
\mathrm{km\, s}^{-1}\mathrm{Mpc}^{-1}$, $\Omega_{\Lambda}=0.7$ and
$\Omega_{\mathrm{M}} = 0.3$.

\section{Observations and Data Reduction}
\subsection{X-Shooter spectroscopy}
\label{xsh-data}
QSO 2222$-$0946 was observed with the X-Shooter spectrograph \citep{Vernet2011} mounted on
ESO's Very Large Telescope, Unit Telescope 2, for a total of 10 hours and 47 minutes during the
nights of 2011 May 14, 2011 August 4, 29, and 30. The integrations were split into ten different
exposures each made up of four sub-exposures dithered along the slit. The instrument is
composed of three separate spectrographs (so-called arms): UVB covering 330~nm to 560~nm,
VIS covering 550~nm to 1020~nm, and NIR covering 1020~nm to 2480~nm. For all
observations slit-widths of 0.8, 0.7 and 0.9 arcsec for were used for 
UVB, VIS and NIR, respectively,
and we used the slow readout with 1$\times$2 binning, i.e., 2 times binning in the spatial direction.
The individual spectra have been processed using the official ESO pipeline
\citep{Modigliani10} for nodded exposures version (2.0.0).
The spectra were individually flux-calibrated using the standard star
observed at the beginning of the night for each integration. All spectra were taken during very
good conditions with clear sky and good seeing. The error on the flux calibration is on
the order of 5\%, as determined from the robustness of the calibration when using different standard stars.
We note that \cite{Fynbo10} scale their spectrum of this target
to the one available from the Sloan Digital Sky Survey. Given our very robust flux calibration we
have chosen to trust the pipeline product. Also, comparison of spectra observed in very different
epochs, as is the case here, is not straightforward given the random variability of QSOs. We encountered
issues with a too low flux-level of the VIS spectra for unknown reasons. For this reason we scaled
the VIS spectra to the overlapping regions of the well-determined UVB and NIR spectra. With a simple
multiplicative factor we were able to match both UVB and NIR at the same time.
The scaling does not affect any of the extracted emission lines as they are all in the NIR
spectra, except for Ly$\alpha$ in the UVB.

The final 2D-spectra were then spatially aligned and added by error-weighting,
and we correct for slit-loss by calculating how much light gets dispersed outside the
slit at the given seeing in each exposure before co-adding all the observations.
We have not corrected for telluric absorption, since those regions are not crucial
for our analysis. The effective seeing in the final flux and wavelength calibrated 2D-spectrum
is 0\,\farcs7 (inferred from the width of the trace at 715~nm) and we
measure the resolving power in the spectrum from telluric lines to be 11,000 and 7,000 for the
VIS and NIR arms, respectively. As there are no direct features in the UVB arm to indicate the
resolving power we assume the resolving power in the UVB arm to be the expected 6,200
from the specifications of the X-Shooter manual given a slit width of 0\,\farcs8, as the average seeing
in UVB was 0\,\farcs77 (measured at 480~nm).

\begin{figure*}
  \includegraphics[width=0.98\textwidth]{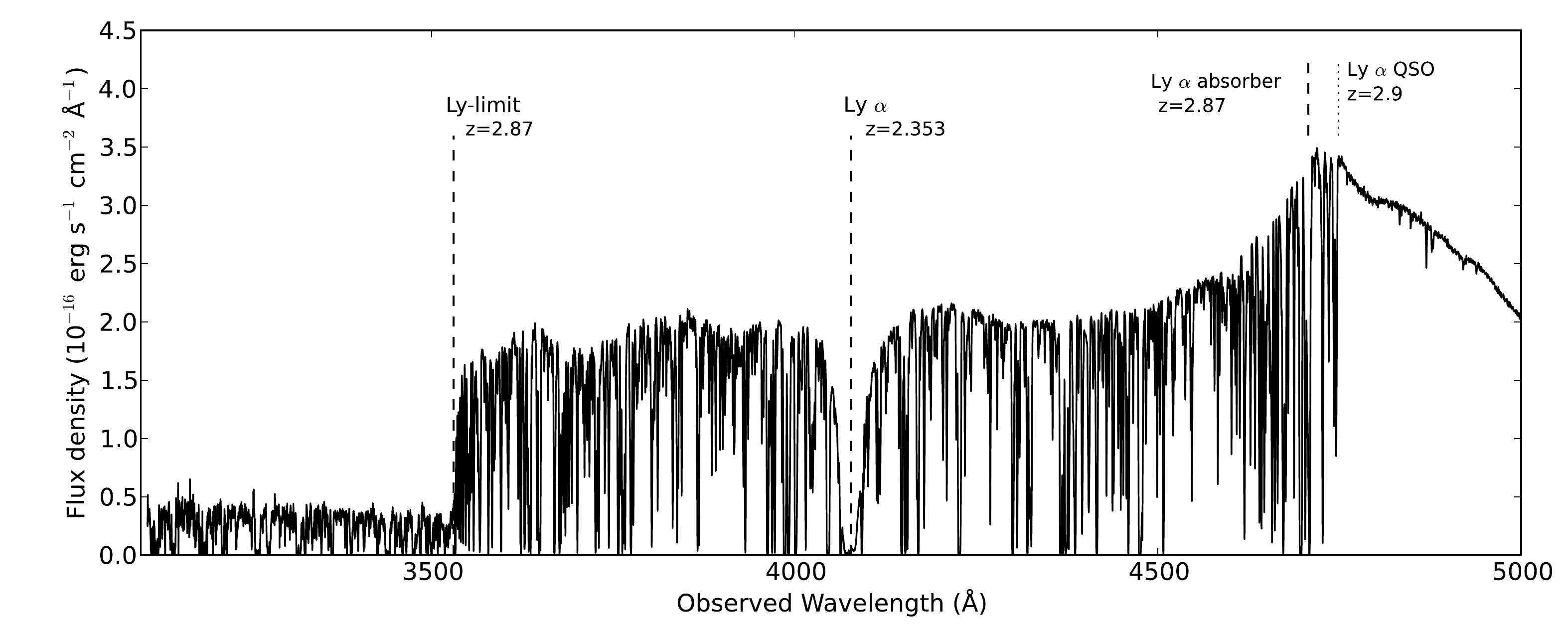}
  \caption{The final 1D spectrum from the UVB arm of the $z=2.9$ Q\,2222$-$0946. In the left
  part of the spectrum, the Lyman limit from an absorber at redshift $z=2.87$ is clearly seen and the
  damped Lyman-alpha absorption line is identified in the middle of the figure by the vertical dashed line.
  The Lyman-alpha emission line at the QSO redshift is seen in the right part of the figure.\label{fig:spec_uvb}}
\end{figure*}

In Figure~\ref{fig:spec_uvb} the spectrum blue-ward of the QSO Ly$\alpha$
emission line is shown, demonstrating the quality of the combined spectrum; note
the significant flux below the Lyman limit at $z=2.87$.

\subsection{\boldmath{$HST$} / WFC3 imaging}
\label{hst-data}

The field was observed with the Wide Field Camera 3 on 2011 Nov 10 (with the UVIS
detector in the F606W filter) and on 2012 Sep 14 and 15 (with the IR detector
in the F105W and F160W filters).  The roll-angle of the telescope was set such
that the galaxy counterpart of the DLA fell between the diffraction
spikes of the Point Spread Function (PSF). The two observations with the IR
detector were taken using the {\small WFC3-IR-DITHER-BOX-MIN} pattern providing an
optimal 4-point sampling of the PSF. The UVIS observation was taken using the
{\small WFC3-UVIS-DITHER-BOX} pattern.

The field was observed using a 4-point sub-pixel dither pattern, which allowed
us to regain more spatial information. We have used the software package
{\tt multidrizzle} \footnote{{\tt multidrizzle} is a product of the Space
Telescope Science Institute, which is operated by AURA for NASA.} to align and
combine the images. By shifting and combining the images taken with sub-pixel
offsets one achieves a better sampling of the PSF, which in the case of the NIR
observations is quite crucial as the PSF is poorly sampled in the native
0\,\farcs13~px$^{-1}$ images. Furthermore, the drizzle-algorithm allows us to
reduce the pixel size when combining the images providing sharper and better
sampled images. For the combination in this work we have set the parameter {\tt
pixfrac} to $0.7$ and used a final pixel scale of 0\,\farcs06~px$^{-1}$ for NIR
and 0\,\farcs024~px$^{-1}$ for UVIS.  For a detailed description of the
parameters in the software we refer the user to the {\tt multidrizzle}$^1$ user
manual.

\section{Results}
\subsection{Spectral PSF subtraction}
In order to detect the faint emission lines from the foreground Damped Lyman-$\alpha$ Absorbing
galaxy we needed to subtract the QSO continuum. We did this by modelling the
spectral trace and the spectral point spread function (SPSF) as a function of wavelength. The trace
position and amplitude were fitted as a function of wavelength in a small (84\AA) region around each
emission line from the galaxy, e.g., $[$O{\sc \,ii}$]$, $[$O{\sc \,iii}$]$, and $[$N{\sc \,ii}$]$.
In the same region around the line we estimated the SPSF by averaging the observed spatial profile
along the spectral direction.
The modelled 2D-spectrum of the QSO was then subtracted from the observed spectrum,
and a 1D-spectrum for each line was extracted with the optimal extraction algorithm of \citet{Horne86}.
All extracted emission lines are shown in Figure~\ref{fig:lines}.

\subsection{Extracting the Emission Lines}
\label{emission}
We were able to detect emission from Ly$\alpha$, H$\alpha$, H$\beta$, the
$[$O{\sc \,ii}$]\,\lambda\lambda \,3727, 3729$ doublet,
the two $[$O{\sc \,iii}$]$ lines at $4959$ \AA\, and $5007$ \AA, and $[$N{\sc \,ii}$]\,\lambda\,6583$.
The line profiles for all the extracted lines were fitted with Gaussian
profiles to measure the flux in the line. The parameters of the Gaussian were
allowed to vary in all fits except for $[$O{\sc \,ii}$]$ and $[$N{\sc \,ii}$]$. The line-width and redshift of the
two $[$O{\sc \,ii}$]$ components were tied in the fit, because of a sky line that falls right between
the two components. For the very faint $[$N{\sc \,ii}$]$-line, we re-binned the spectrum by a factor of
two and fixed the redshift and line-width to the values of the nearby H$\alpha$ line.
The errors on the fluxes were estimated by varying the line profile within the error of each fit parameter
1,000 times. The uncertainty was then determined from the 1\,$\sigma$ width of the resulting distribution
of fluxes. The fluxes and errors are listed in Table~\ref{FluxTable}.

\begin{figure}
  \includegraphics[width=0.49\textwidth]{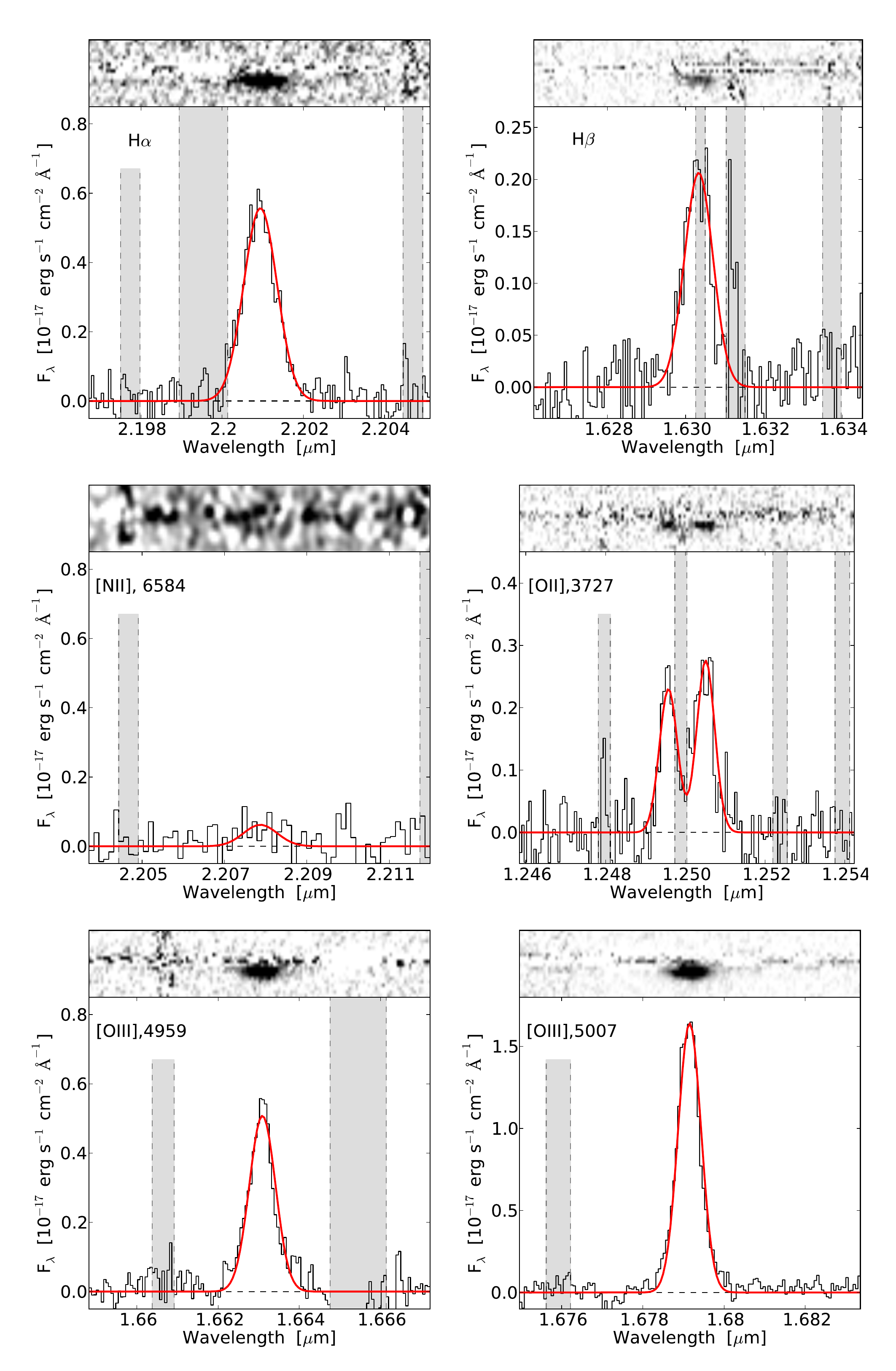}
  \caption{Emission lines extracted after the subtraction of the QSO continuum in the 2D spectrum. Each panel
  shows the 2D spectrum where the horizontal residuals from the QSO subtraction is seen ({\it top})
  and the extracted 1D spectrum of the line ({\it bottom}). The wavelengths are given in $\mu$m and all
  flux-density units are $10^{-17}~\mathrm{erg}~\mathrm{s}^{-1}~\mathrm{cm}^{-2}~\mathrm{\AA}^{-1}$.
  The red line in each panel indicates the best fitting Gaussian to the line profile. The grey filled
  areas indicate skylines or telluric absorption features.
   \label{fig:lines}}
\end{figure}

\begin{figure}
  \includegraphics[width=0.48\textwidth]{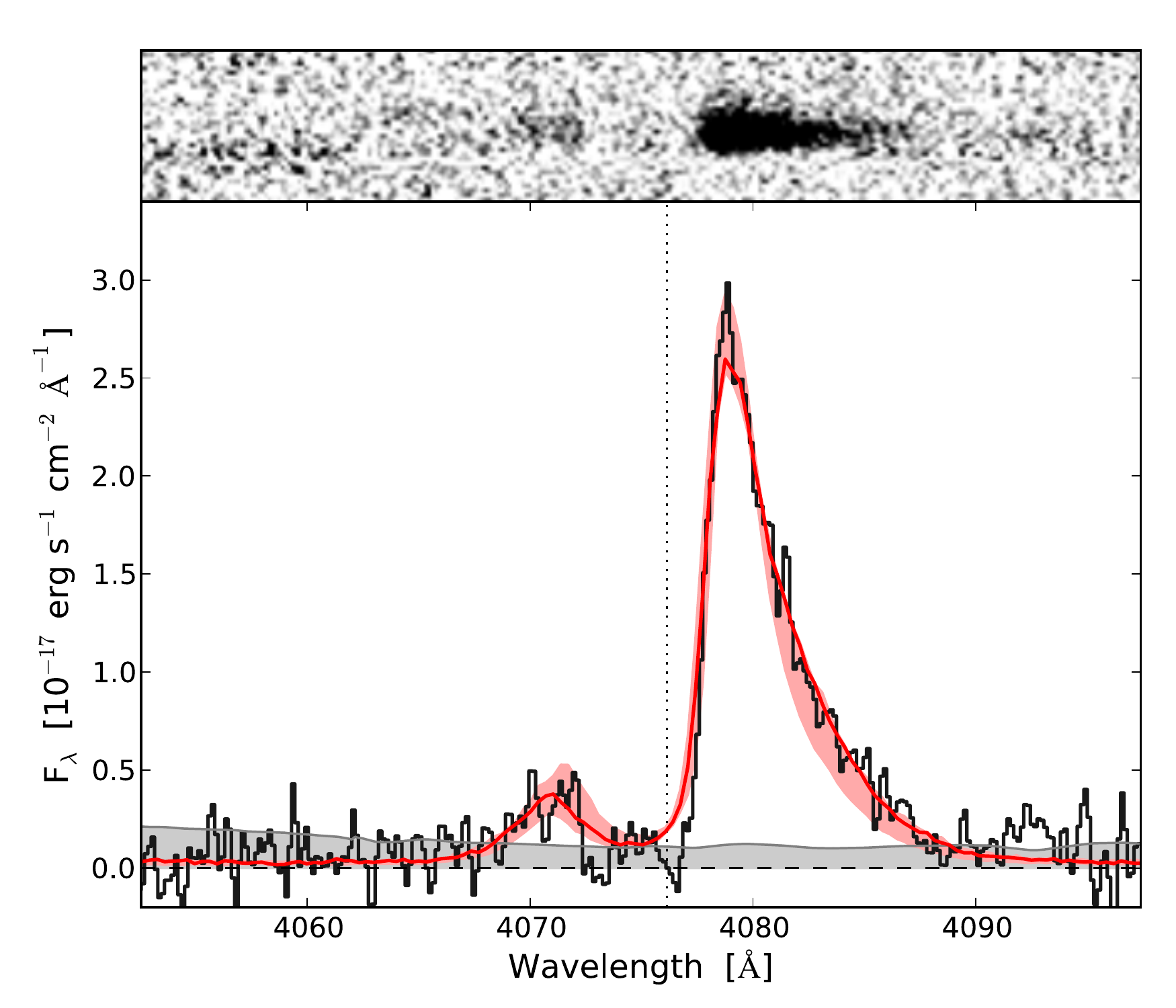}
  \caption{The Ly$\alpha$ emission line extracted from the DLA trough in the X-shooter spectrum.
  The top panel shows the observed 2D spectrum and the panel below shows the extracted
  1D spectrum, and 1$\sigma$ uncertainty (grey shaded area). The dotted vertical line shows the systemic
  line centre. The common asymmetric line shape 
  of Ly$\alpha$ is seen in this case, where the blue-shifted part of the line is heavily absorbed.
  On top of the observed spectrum the best fitting model of the emission profile is shown (red line)
  along with the 68\% confidence interval (red shaded area), see Sect.~\ref{lya-model} for details.
   \label{fig:lyalpha}}
\end{figure}

All the extracted lines with Gaussian line fits are shown in Fig.~\ref{fig:lines} along with the best fitting
Gaussian profile. The grey shaded areas indicate telluric emission or absorption features. These regions were
included in the fits to give the optimal estimate of uncertainties on the fit parameters. The Ly$\alpha$
line is shown in Fig.~\ref{fig:lyalpha}. Due to the asymmetric shape and the complex nature of the
resonant line, we determine the line flux simply by integrating the observed line profile as opposed to
fitting the line.

\begin{table}
\caption{Measured emission line fluxes\label{FluxTable}}
\begin{center}
\begin{tabular}{l c r c c}
\hline
\hline
Transition & Wavelength$^{(1)}$ & Flux$^{(2)}$ & FWHM$^{(3)}$ & $z$ \\
\hline
Ly$\alpha$	&	1215.67	&	14.3$\pm$0.3 & -- & -- \\
$[$O{\sc \,ii}$]$	&	3726.03,\,3728.82	&	2.9$\pm$0.3 & $121\pm 8$~~ & 2.3536\\
H$\beta$		&	4861.33	&	1.9$\pm$0.2 & $151\pm15$ & 2.3537 \\
$[$O{\sc \,iii}$]$	&	4958.92	&	4.1$\pm$0.3 & $129\pm 5$~~ & 2.3537 \\
$[$O{\sc \,iii}$]$	&	5006.84	&	11.6$\pm$0.6 & $110\pm 2$~~ & 2.3537 \\
H$\alpha$	&	6562.80	&	5.7$\pm$0.3 & $124\pm 4$~~ & 2.3537 \\
$[$N{\sc \,ii}$]$	&	6583.41	&	0.6$\pm$0.2 & -- & -- \\
\hline
\end{tabular}
\end{center}

$^{(1)}$ Transition rest frame wavelength in $\mathrm{\AA}$.\\
$^{(2)}$ Flux in units of $10^{-17}$erg s$^{-1}$ cm$^{-2}$, before reddening correction.\\
$^{(3)}$ Line width at FWHM in units of km~s$^{-1}$ corrected for the instrumental resolution of 45~km~s$^{-1}$.

\end{table}

\subsection{Balmer Decrement}
\label{balmer}
The ratio of the H$\alpha$ and H$\beta$ line fluxes provides us with information
about the dust extinction in the system as we know what this ratio should be intrinsically, given
the physical conditions of the emitting region. Assuming case B recombination with an
electron temperature of $T_e = 10^4$~K and density of $n_e=10^2$~cm$^{-3}$, which are
standard assumptions in the literature for star forming regions, the line ratio has a value of $2.86$.
Requiring that our measured line ratio, after reddening, and the intrinsic ratio be the same, we
estimate the reddening:
$$ E(B-V) = \frac{2.5}{k(\mathrm{H}\beta)-k(\mathrm{H}\alpha)}
\log\left(\frac{(\mathrm{H}\alpha/\mathrm{H}\beta)_{\mathrm{obs}}}
                      {(\mathrm{H}\alpha/\mathrm{H}\beta)_{\mathrm{0}}}\right) , $$\\
where $k$ denotes the reddening law evaluated at the given wavelength and
$(\mathrm{H}\alpha/\mathrm{H}\beta)_{\mathrm{0}}$ indicates the intrinsic line ratio.
We use the extinction law from \citet{Calzetti2000} to quantify
the extinction while adopting a $R_V$ value of 4.05. From the measured ratio in our spectrum of
$(\mathrm{H}\alpha\,/\,\mathrm{H}\beta)_{\mathrm{obs}} = 3.03\pm0.34$ we obtain the
corresponding extinction of $E(B-V)$ = 0.05$\pm$0.10. In this calculation we have not
taken into account effects from differential slit-loss,
which arise from the wavelength dependence of the seeing and atmospheric dispersion.
The slit loss is greater at lower wavelengths, i.e., for the H$\beta$ line in the case of the Balmer line ratio.
However, the effect in the $K$- and $H$-band, where we extract the Balmer lines, is minor ($\sim2$\%)
compared to the uncertainty introduced by the sky lines, especially at the position of the H$\beta$-line.

\subsection{Metallicity}
We have inferred the metallicity of the system using three independent methods;
two measures of the emission-line metallicity using the strong line ratios R$_{23}$ and N2, respectively,
and one measure of the absorption-line metallicity from Voigt-profile fitting. In the next sections we present
each determination in detail.

\subsubsection{R$_{23}$ calibration}
We determined the metallicity $12+\log$(O/H) by use of the strong line diagnostic, R$_{23}$ \citep{Pagel1979},
defined as the ratio ($[$O{\sc \,ii}$]\,\lambda\lambda\,3727,3729 + [$O{\sc \,iii}$]~\lambda \,4959 + [$O{\sc \,iii}$]~\lambda \,5007$) / H$\beta$.
One complication of using this diagnostic is that the metallicity is double-valued for a given
value of R$_{23}$ (the so-called {\it upper} and {\it lower} branches). However, due to our
relatively high value of R$_{23}$ we are in the region of the diagram where the two branches
are close to one another. The calibration by \citet{Kobulnicky2004} that we use depends
on the ionization parameter, $q$, which can be determined by using the line ratio
O$_{32} = [$O{\sc \,iii}$]\, \lambda 5007$~/~$[$O{\sc \,ii}$]\, \lambda3727$, but requires prior
knowledge of the metallicity. Given the two measured quantities, R$_{23}$ and O$_{32}$, we can
solve for both metallicity and $q$ at the same time by iterating the two expressions given an initial
guess of metallicity. 

\begin{figure}
  \includegraphics[width=0.48\textwidth]{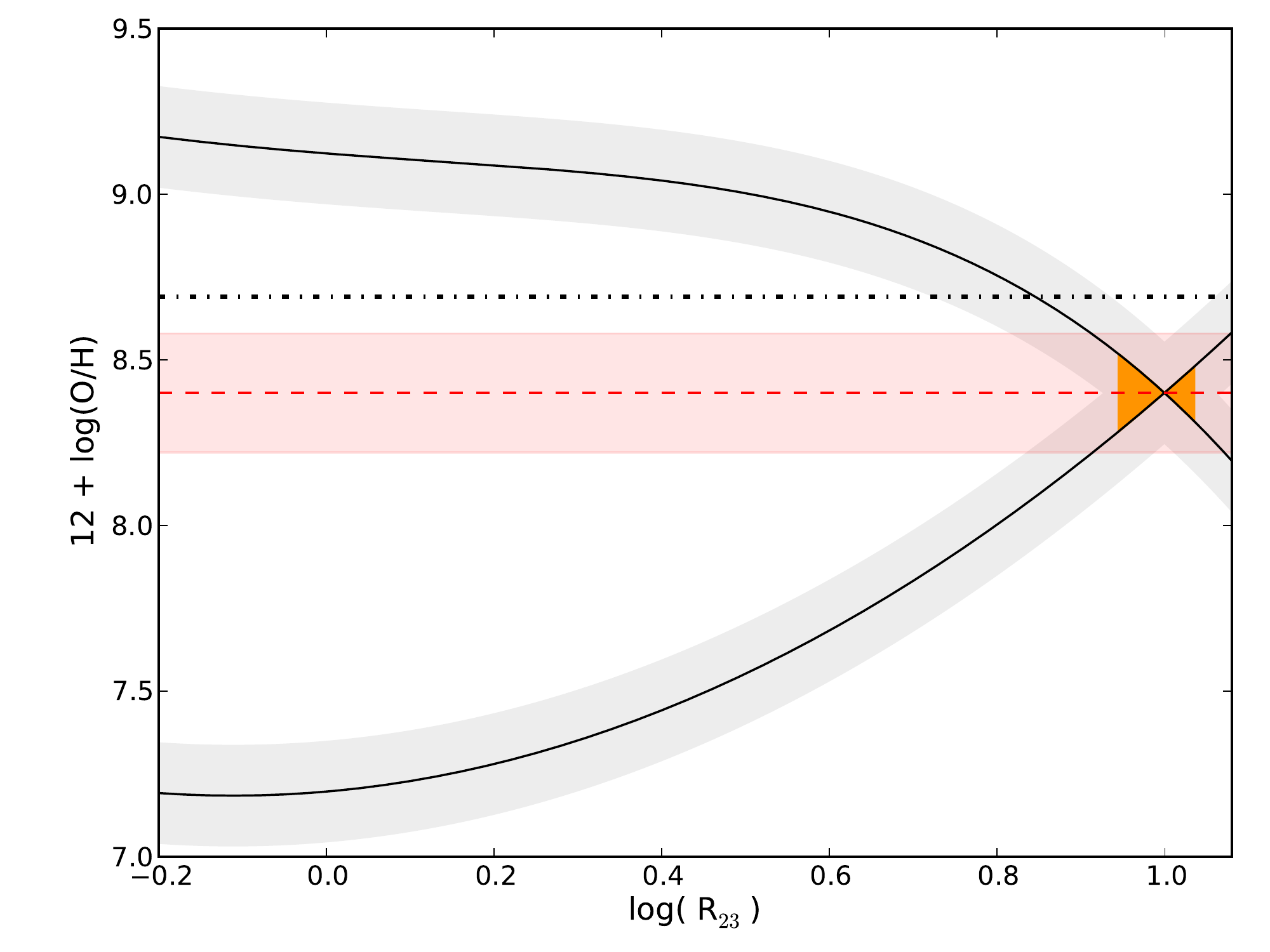}
  \caption{Plot of the R$_{23}$ strong line ratio using the calibration by \citet{Kobulnicky2004} (two solid lines)
  with model uncertainty of $0.15$~dex indicated by the light grey shaded area around the curves.
  The orange filled area between the two solid lines indicates the constraint from our measurement of R$_{23}$.
  The dashed line indicates the inferred metallicity of $12+\log(\mathrm{O/H})=8.41$ and the red shaded area
  around it indicates the $1\sigma$ uncertainty of $0.19$~dex. The dot-dashed line shows the Solar oxygen
  abundance of $12+\log(\mathrm{O/H})_{\odot}=8.69$ \citep{Asplund2009}.
   \label{fig:R23}}
\end{figure}

From our spectrum we measure log(R$_{23})=1.00\pm0.05$ and log(O$_{32})=0.55\pm0.05$. These
line ratios are corrected for reddening as indicated by the Balmer decrement. However,
in Sect.~\ref{sfr} we infer $E(B-V)=0.06\pm0.01$, which is the value that we have adopted
for the correction here. We perform the iterative calculation of metallicity 500 times while
varying the two line-ratios (R$_{23}$ and O$_{32}$) within their respective errors. The metallicity
and uncertainty is then determined as the median and standard deviation of the resulting distribution
This in turn yields an inferred metallicity of $12+\log(\mathrm{O/H})=8.41\pm0.19$, see
Fig.~\ref{fig:R23}. We furthermore note that the result does not depend on the initial guess,
and that the model uncertainty introduced by the error on the ionization parameter, $q$, is negligible.
The uncertainty of $0.19$~dex includes the contribution from the scatter in the R$_{23}$ calibration
which we conservatively set to $0.15$~dex.

\subsubsection{N2 calibration}
Our detection of $[$N{\sc \,ii}$]$ enables us to infer the metallicity independently
from the N2 index (\,$[$N{\sc \,ii}$]$ / H$\alpha$\,) as
this line ratio has been found to correlate with metallicity \citep*{Pettini2004}. The calibration from
these authors give $12+\log(\mathrm{O/H}) = 8.90 + 0.57\cdot \log(\mathrm{N2})$
with a scatter of $\pm$0.18 dex.
The small wavelength difference of H$\alpha$ and $[$N{\sc \,ii}$]$ means that we do not need
to worry about reddening effects in the N2 ratio. The measured line fluxes yield a line ratio of
N2 $=0.11\pm0.03$, giving an inferred metallicity of $12+\log(\mathrm{O/H}) = 8.36\pm0.19$.\\
\citet{Peroux12} infer an upper limit of $12+\log({\rm O/H}) < 8.2$ from their non-detection of
[\ion{N}{ii}]. Our measurement is slightly higher, but still consistent with theirs within the uncertainties.

Since the two measures of the emission metallicity, R$_{23}$ and N2, are independent we
can combine both measures to obtain a more precise estimate of
$12+\log(\mathrm{O/H}) = 8.39\pm0.13$. Adopting the Solar abundance presented by
\citet{Asplund2009} this gives us $[\mathrm{O/H}]=-0.30\pm0.13$.

\subsection{Absorption-line Abundances}
\label{absfit}
In the following section we present detailed measurements of the absorption-line
abundances detected in the DLA system.
\citet{Fynbo10} present an analysis of the metal absorption-line system of this
DLA using a three-component Voigt-profile fit. They derive fairly high metallicities from
$[$Zn/H$]=-0.46\pm0.07$, $[$Si/H$]=-0.51\pm0.06$, and $[$S/H$]=-0.6\pm0.1$, and
see evidence for dust depletion from $[$Mn/H$]=-1.23\pm0.06$, $[$Ni/H$]=-0.85\pm0.06$,
and $[$Fe/H$]=-0.99\pm0.06$.
Since our data have much better resolution (recall values in Section~\ref{xsh-data}
compared to $R=4700,6700,4400$ in the Fynbo et al. study for the UVB, VIS and NIR arms, respectively)
and much higher Signal-to-Noise Ratio (SNR$\sim$100, compared to their SNR$\sim$50) we can
alleviate the effect of hidden saturation, thereby further constraining the metal abundances.

\begin{figure}
  \includegraphics[width=0.48\textwidth]{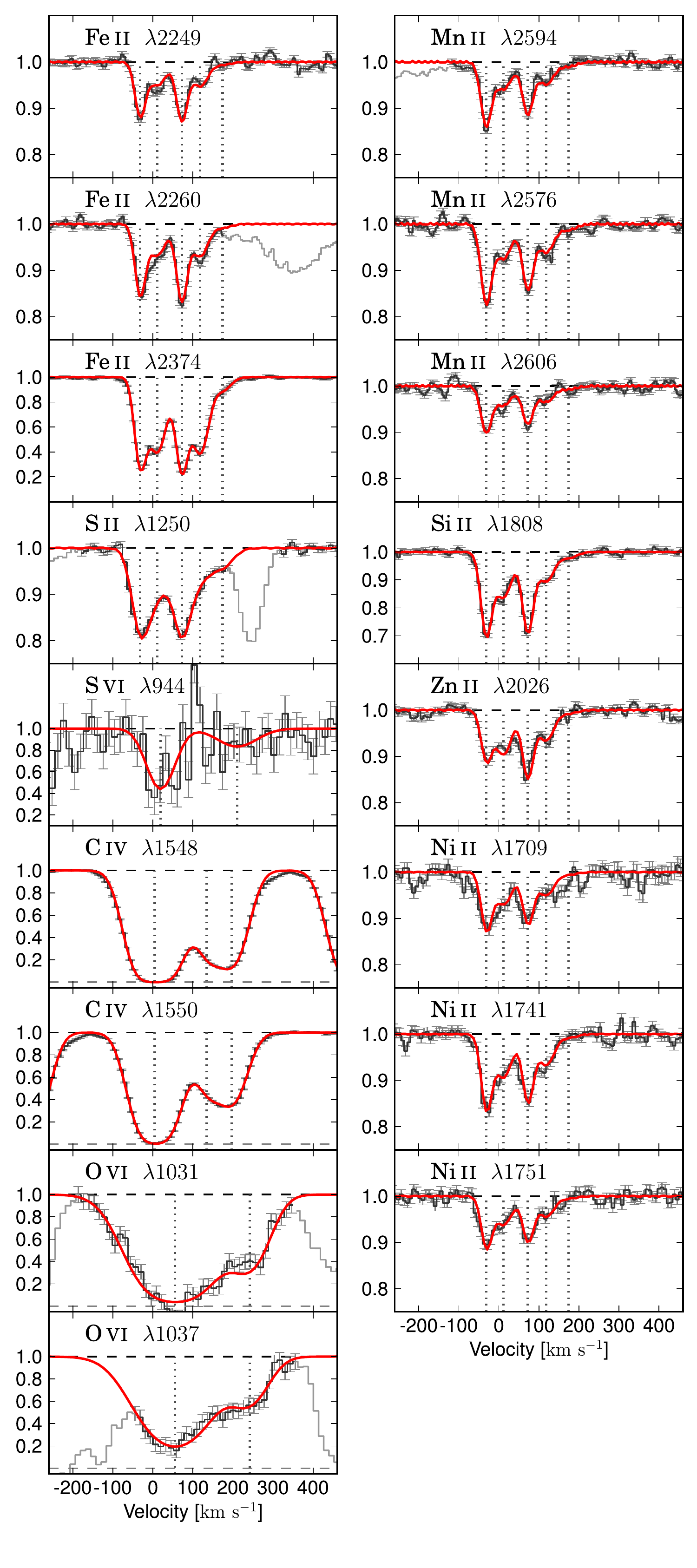}
  \caption{Results of Voigt-profile fitting to the metal absorption lines.
  The zero point of the velocity scale is fixed to the redshift from the emission lines,
  $z_{\rm em}=2.3537$ with an uncertainty of 9 \kms. The figure shows the normalized
  spectra around each fitted line. The part of the spectrum plotted in light grey without error-bars
  in the panels of \ion{Mn}{ii}~$\lambda$2594, \ion{S}{ii}~$\lambda$1250,
  \ion{Fe}{ii}~$\lambda$2260, and \ion{O}{vi}~$\lambda\lambda$1031,1037
  shows absorption unrelated to the given line.
  \label{fig:metals}}
\end{figure}

In order to properly decompose the absorption profiles in different velocity
components we have selected unsaturated, but well-defined low-ionization lines
to infer the metallicity of the absorbing gas (see Fig.~\ref{fig:metals}). Furthermore,
we have access to a few high-ionization lines: \ion{C}{iv}~$\lambda\lambda$\,1548, 1550,
\ion{O}{vi}~$\lambda\lambda$\,1031, 1037, and possibly \ion{S}{vi}~$\lambda$944.
We have ignored the Cr{\sc \,ii}~$\lambda$\,2066 line because it is heavily
blended with telluric absorption features.
We found that a five-component fit provides a satisfactory description of the low-ionization lines.
All the lines were fitted simultaneously using {\tt FitLyman} in {\small MIDAS} \citep{FitLyman}
while tying the broadening parameter, $b$, and the
redshift for each velocity component for all lines. The results for the individual components
of the fit are summarized in Table~\ref{MetalTable}.
We derive the total column densities from the fit by adding the column densities for
all components of each species. These and the derived metallicities are
listed in Table~\ref{tab:metal}.
For the Ti{\sc \,ii}~$\lambda\lambda$\,1910 doublet we get an upper limit (3$\sigma$) of
$[$Ti/H$]\le-0.1$, as the line is shallow and
the continuum level in this part of the spectrum is uncertain.
The abundance of Zn is probably overestimated due to blending with the weak
\ion{Mg}{i}~$\lambda$\,2026 line. However, from the \ion{Mg}{i}~$\lambda$\,2852 line
we are able to constrain the amount of contamination from \ion{Mg}{i} to be $\sim$0.1 dex.

The inferred metallicities are in very good agreement with those determined
by \citet{Fynbo10}. The uncertainty of our metallicities, which we estimate to be
0.05 dex from the variance in the data, is dominated by the uncertainty on the
normalization of the QSO continuum.

For the high-ionization lines, we have let the redshifts and broadening parameters
vary freely for each species. We used three components to fit \ion{C}{iv} whereas
we used two components for the more noisy \ion{O}{vi} and \ion{S}{vi} lines.
The results from the fit are listed in the bottom part of Table~\ref{MetalTable}.
Our possible detection of \ion{S}{vi} is quoted as an upper limit, and the first component
of \ion{C}{iv} is quoted as a lower limit since the line is saturated. The velocity of the
main component of \ion{C}{iv} is fully consistent with the systemic velocity from
the emission lines within the uncertainty of 9 \kms. The oxygen and sulfur lines
have slightly larger offsets relative to the emission redshift of +20 \kms and +55 \kms, respectively.

\begin{table}
\caption{Ionic column densities for individual absorption components of low-ionization lines (top)
and high-ionization lines (bottom).
The errors quoted below only include the formal errors from {\tt FitLyman}.\label{MetalTable}}
\begin{center}
\begin{tabular}{l c c}
\hline
\hline
Component $^1$	&	$\log(N/ {\rm cm}^{-2})$	&	$b$ \\
		&	 			&	km~s$^{-1}$ \\
\hline
Si{\sc \,ii}~~$\lambda$\,1808 & & \\
$v=-32$ \kms&  15.11 $\pm$ 0.01 & 9.1 $\pm$ 0.1   \\
$v=11 $ \kms&  14.97 $\pm$ 0.01 & 24.0 $\pm$ 0.2  \\
$v=73 $ \kms&  15.10 $\pm$ 0.01 & 10.1 $\pm$ 0.1  \\
$v=118$ \kms&  14.73 $\pm$ 0.02 & 21.6 $\pm$ 0.3  \\
$v=174$ \kms&  14.05 $\pm$ 0.07 & 23.7 $\pm$ 1.5  \\

 & & \\
S{\sc \,ii}~~$\lambda$\,1250 & & \\
$v=-32$ \kms&  14.80 $\pm$ 0.01 & 9.1 $\pm$ 0.1   \\
$v=11 $ \kms&  14.56 $\pm$ 0.02 & 24.0 $\pm$ 0.2   \\
$v=73 $ \kms&  14.82 $\pm$ 0.01 & 10.1 $\pm$ 0.1   \\
$v=118$ \kms&  14.35 $\pm$ 0.03 & 21.6 $\pm$ 0.3  \\
$v=174$ \kms&  14.18 $\pm$ 0.04 & 23.7 $\pm$ 1.5  \\

 & & \\
Mn{\sc \,ii}~~$\lambda\lambda\lambda$\,2576, 2594, 2606 & & \\
$v=-32$ \kms&  12.41 $\pm$ 0.01 & 9.1 $\pm$ 0.1   \\
$v=11 $ \kms&  12.21 $\pm$ 0.01 & 24.0 $\pm$ 0.2  \\
$v=73 $ \kms&  12.33 $\pm$ 0.01 & 10.1 $\pm$ 0.1  \\
$v=118$ \kms&  12.09 $\pm$ 0.01 & 21.6 $\pm$ 0.3  \\
$v=174$ \kms&  11.43 $\pm$ 0.05 & 23.7 $\pm$ 1.5  \\

 & & \\
Fe{\sc \,ii}~~$\lambda\lambda\lambda$\,2249, 2260, 2374 & & \\
$v=-32$ \kms&  14.59 $\pm$ 0.01 & 9.1 $\pm$ 0.1	  \\
$v=11 $ \kms&  14.39 $\pm$ 0.01 & 24.0 $\pm$ 0.2  \\
$v=73 $ \kms&  14.64 $\pm$ 0.01 & 10.1 $\pm$ 0.1  \\
$v=118$ \kms&  14.38 $\pm$ 0.01 & 21.6 $\pm$ 0.3  \\
$v=174$ \kms&  13.46 $\pm$ 0.02 & 23.7 $\pm$ 1.5  \\

 & & \\
 Ni{\sc \,ii}~~$\lambda\lambda\lambda$\,1709, 1741, 1751	& & \\
$v=-32$ \kms&  13.49 $\pm$ 0.01 & 9.1 $\pm$ 0.1   \\
$v=11 $ \kms&  13.41 $\pm$ 0.02 & 24.0 $\pm$ 0.2  \\
$v=73 $ \kms&  13.45 $\pm$ 0.01 & 10.1 $\pm$ 0.1  \\
$v=118$ \kms&  13.23 $\pm$ 0.02 & 21.6 $\pm$ 0.3  \\
$v=174$ \kms&  12.31 $\pm$ 0.21 & 23.7 $\pm$ 1.5  \\

 & & \\
Zn{\sc \,ii}~~$\lambda$\,2026 & & \\
$v=-32$ \kms&  12.14 $\pm$ 0.01 & 9.1 $\pm$ 0.1   \\
$v=11 $ \kms&   12.29 $\pm$ 0.01 & 24.0 $\pm$ 0.2  \\
$v=73 $ \kms&   12.31 $\pm$ 0.01 & 10.1 $\pm$ 0.1  \\
$v=118$ \kms&  12.08 $\pm$ 0.02 & 21.6 $\pm$ 0.3  \\
$v=174$ \kms&  11.16 $\pm$ 0.15 & 23.7 $\pm$ 1.5  \\

 & & \\
\hline
\hline
 & & \\
C{\sc \,iv}~~$\lambda\lambda$\,1548, 1550 & & \\
$v=5$ \kms 	&  $>$15.23 $^2$ &	46.7 $\pm$ 0.3	\\
$v=134$ \kms 	&  14.11 $\pm$ 0.02 &	28.0 $\pm$ 1.4	\\
\vspace{0.2cm}
$v=197$ \kms	&  14.34 $\pm$ 0.01 &	36.5 $\pm$ 0.8	\\
O{\sc \,vi}~~$\lambda\lambda$\,1031, 1037 & & \\
$v=55$ \kms 	&  15.24 $\pm$ 0.02 &	105 $\pm$ 6	\\
\vspace{0.2cm}
$v=242$ \kms 	&  14.49 $\pm$ 0.08 &	54 $\pm$ 10	\\
S{\sc \,vi}~~$\lambda$\,944 & & \\
$v=20$ \kms 	& $\le$14.1 $\pm$ 0.1 $^3$ &	35 $\pm$ 9	\\
$v=210$ \kms	& $\le$13.6 $\pm$ 0.2 $^3$ &	61 $\pm$ 33	\\

\hline

\end{tabular}
\end{center}
$^{1}$ The velocity, $v$, indicated at each component shows the relative\\
velocity with respect to the emission redshift $z=2.3537$.\\
$^{2}$ Lower limit due to saturation.\\
$^{3}$ Upper limit due to possible blending.
\end{table}

\begin{table}
\caption{Total column densities and metallicities for the low-ionization lines in the $z=2.354$
absorbing system.
\label{tab:metal}}
\begin{center}
\begin{tabular}{l c c c}
\hline
\hline
Element	&	$\log(N_X / {\rm cm}^{-2})$	&	[X/H]	&	$\log(N_{\odot}/ {\rm cm}^{-2})$	\\
		&			&		&	\\
\hline
H	&	20.65 $\pm$ 0.05	& --	&	12.00	\\
Si	&	15.62 $\pm$ 0.01	&  $-$0.54 $\pm$ 0.05  & 7.51 \\
S	&	15.31 $\pm$ 0.01	&  $-$0.49 $\pm$ 0.05  & 7.15 \\
Mn	&	12.89 $\pm$ 0.01	&  $-$1.19 $\pm$ 0.05  & 5.43 \\
Fe	&	15.13 $\pm$ 0.01	&  $-$1.02 $\pm$ 0.05  & 7.50 \\
Ni	&	14.02 $\pm$ 0.01	&  $-$0.85 $\pm$ 0.05  & 6.22 \\
Zn	&	12.83 $\pm$ 0.01	&  $-$0.38 $\pm$ 0.05  & 4.56 \\
\hline
\end{tabular}
\end{center}
Solar abundances are taken from \citet{Asplund2009}.
\end{table}

\subsubsection{Molecular hydrogen}
\label{H2}
Our data exhibit a number of dips at the expected position of H$_2$ lines at $z_{\rm abs}=2.354$,
which suggest a detection of molecular hydrogen at the level of
$\log(N_{{\rm H}_2} / {\rm cm}^{-2})\sim 14$.
However, asserting the presence of H$_2$ and deriving accurate column densities or limits
requires much higher spectral resolution data than what we have available here
\citep[see, e.g.,][]{Ledoux03,Noterdaeme08}.
This is needed in order to ({\it i}) deblend the possible H$_2$ lines from the Lyman-$\alpha$
forest, ({\it ii}) estimate the true continuum of the QSO from unabsorbed spectral regions,
and ({\it iii}) resolve the velocity structure of the system. \citet{Fynbo11} reported the detection
of H$_2$ lines at a resolving power of $R=6400$ in the $z_{\rm abs}=2.58$ DLA towards SDSS
J\,0918$+$1636, but the absence of strong damping wings makes the column density
estimates uncertain ($\log N($H$_2)\sim 16 -19$). The present tentative detection of H$_2$ lines
towards SDSS J\,2222$-$0946 and the corresponding column densities should therefore be verified
via follow-up high resolution spectroscopy (i.e., $R\ga 50,000$).

\subsection{Structural Fitting of the \boldmath{$HST$} images}
\label{galfit}
We started out by modelling the Point Spread Function in each image using
the software {\sc TinyTim} to create sub-sampled PSFs
simulated for each of the individual frames at each position
in the four-point dither pattern. We preferred this approach over using stellar sources as
the {\sc TinyTim} PSFs are more sensitive to the outer regions of the PSF, where the
stellar sources are dominated by noise in the sky background.\\
After resampling the modelled PSFs to
the native pixel size and convolving with the filter-specific Charge Diffusion Kernel, these "raw" PSF images
were drizzled together in the same way as the actual data to replicate the effect of {\tt multidrizzle} on the
PSF shape. We were not able to simulate the PSFs for the infrared images properly because the
images were mildly saturated. We therefore chose to use stellar PSFs generated by median
combination of stars in the field. Since there are only a handful isolated stars in the field of view
with signal-to-noise ratios similar to that of the QSO, we used a few stars with slightly lower
SNR in order to reduce the background noise in the PSF.

We used the software {\sc Galfit} \citep{Peng02} to subtract the QSO and to characterize the absorbing galaxy.
Using the modelled PSF for the F606W image we first subtracted the QSO by simply modelling it as
a point source while modelling a constant sky background. We then located the nearby galaxy in
the residuals. Hereafter, we re-did the fit, this time simultaneously fitting the QSO, the background, and
the galaxy using a S\'ersic surface brightness profile:
$$ \Sigma(r) = \Sigma_e\,\exp \left[ -\kappa \left( \left(\frac{r}{r_e}\right)^{1/n} -1 \right) \right] ,$$
\\
where $\Sigma_e$ is the surface brightness at the effective radius, $r_e$, defined as the radius enclosing
half the flux. The S\'ersic index, $n$, determines the concentration of the profile, with high $n$ profiles having
steeper inner slopes and larger extended wings. The opposite is the case for low $n$. The parameter $\kappa$
is linked to $n$ to ensure that half of the light is enclosed within the effective radius.
The software G{\small ALFIT} uses a 2D S\'ersic profile allowing for elliptical isophotes.
The output from fitting the S\'ersic profile is given in terms of the total flux from the integrated profile,
the effective semi-major axis, $a_e$, the S\'ersic index, $n$, the axis ratio of semi-major and -minor axes,
$b/a$, and the angle of $a_e$ with respect to the image-axes.
We fitted the galaxy allowing all parameters to vary freely. This resulted in the following best-fitting
parameters: $\mathrm{mag_{AB}}=24.29\pm0.04$, $n=0.95\pm0.15$, $a_e=5.80\pm0.30$~px,
$b/a=0.17\pm0.02$, and $\mathrm{PA}=-26.45\pm1.34^{\mathrm{o}}$ in the F606W image.

Due to the broader PSF in the NIR images the QSO light is spatially overlapping with the galaxy.
We therefore fixed the structural parameters in the S\'ersic fit of the galaxy to those of the
well-constrained F606W fit. We note that simply locking all structural parameters of the
S\'ersic profile in the NIR images might not give the most accurate description of the galaxy,
as parameters such as size and $n$ depend on wavelength and thus yield different results
when analysed in different wavelength band-passes \citep{Kelvin2012}. However, in order to
obtain the most reliable fluxes and to get the fit to converge we had to keep the variables fixed.
In order to estimate how robust our obtained fluxes are with respect to the parameters that were
held fixed, we varied the S\'ersic parameters within their errors (as given by the fit to F606W)
and re-did the fit for each new set of profile parameters. The uncertainty on the flux was very
minor (0.03~dex) compared to the large uncertainty caused by the PSF subtraction ($\sim$0.2~dex).
All the obtained magnitudes are listed in Table~\ref{GalfitTable}.

\begin{figure}
  \includegraphics[width=0.48\textwidth]{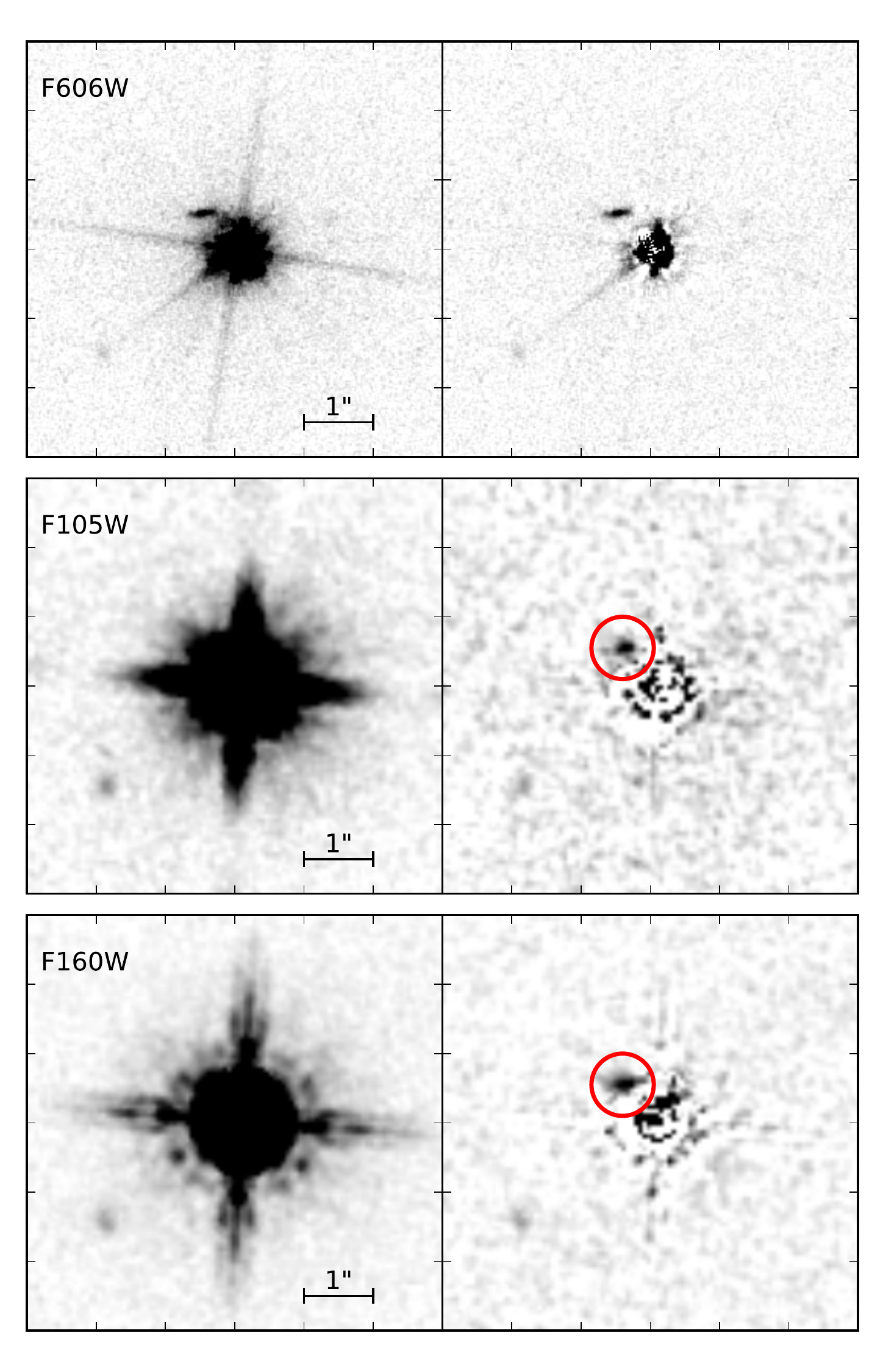}
  \caption{Data images ({\it left}) and PSF-subtracted residuals ({\it right}) from {\sc Galfit} in
  all three filters. All images are aligned North up and East left. The galaxy causing the absorption
  is clearly visible in the F606W image even before PSF subtraction. The same position is shown by the red
  circle in the IR images.
   \label{fig:galfit}}
\end{figure}

In Fig.~\ref{fig:galfit} we show the WFC3 images in the three filters and the residuals
after subtracting the modelled QSO PSF. All images are rotated to have North up and East left.
The galaxy causing the DLA seen in the QSO spectrum is located to the North-East of the QSO.
The red circle in the F105W and F160W images indicates the position of the galaxy from the F606W image.
The identification is based on the fact that the position is consistent with the measured offset between the
QSO trace and the emission lines in the X-shooter spectrum. The impact parameter of the galaxy with
respect to the background quasar is 0\,\farcs74 corresponding to a projected distance of 6.3~kpc at
$z=2.354$, and the angle between the major axis of the galaxy and the line connecting the QSO to
the central region of the galaxy is $71\pm2^{\mathrm{o}}$.

\begin{table}
\caption{Results from {\sc Galfit} analysis\label{GalfitTable}}
\begin{center}
\begin{tabular}{l c c c c c}
\hline
\hline
FILTER & mag~$[${\sc ab}$]$ & $a_e\,/$~kpc  &	$n$	&	$b/a$ \\
\hline
F606W	&	24.29 $\pm$ 0.04	&	$1.12\pm0.06$ &	$0.95\pm0.15$ & $0.17\pm0.02$  \\
F105W	&	24.51 $\pm$ 0.21	&	$1.12 ^*$ & $0.95^*$ & $0.17^*$ \\
F160W	&	23.53 $\pm$ 0.13	&	$1.12 ^*$	& $0.95^*$ & $0.17^*$ \\
\hline
\end{tabular}
\end{center}

$^*$ Parameters that were fixed in the corresponding fit.\\

\end{table}

\subsection{Ly$\alpha$ emission modeling}
\label{lya-model}
We observe a typical, double-peaked \lya\ emission line, with a strong
component redwards of line centre and a less prominent blue component
(Fig.~\ref{fig:lyalpha}). The large difference between the two peaks
indicates that the \lya\ photons escape through an expanding medium,
and the purpose of this section is to investigate what constraints we can
put on the outflow velocity, \Vout. Qualitatively we can see that the red wing extends
to several hundreds of \kms. However, in order to get a more accurate
description of the emission profile we have constructed a semi-realistic model
of the system and run \lya\ radiative transfer (RT) through it, to find the
best-fitting spectrum. The RT is conducted using the code {\sc MoCaLaTA}
\citep{Laursen09a,Laursen09b}, while the galaxy model is similar to the one
described in \citep{Laursen2013}. In short, the galaxy is modelled as a sphere
of multiphase gas, with warm, neutral clouds floating in a hot, predominantly
ionized medium. A similar procedure was used in
\citet{Noterdaeme12}, although here we employ a more rigorous approach.

A high number of parameters dictate the outcome of such a simulation. Luckily,
our observations offer excellent constraints on many of these: The effective
radius, $r_e = 1.12$~kpc, used to model the size of the emitting region,
the metallicity $Z \simeq 0.31 \Zsun$ (we assume that
the amount of metals that condense to dust is similar to the local Universe
\citep{Zafar2013}), and an average of 130 \kms\ from
the measured emission-line widths are used as a proxy for the
intrinsic \lya\ line width. From the widths of the low-ionization absorption
lines, we use a velocity dispersion of the clouds of 115 \kms\ . Finally, we infer
an intrinsic equivalent width of 150 {\AA}, in accordance with the F606W magnitude.
For details on the rest of the parameters, e.g., cloud size distribution,
temperatures and densities of the two phases, etc., see \citet{Laursen2013}.

We set $r_{\mathrm{gal}} = 10$~kpc, but note that the
exact size of the system is not important, rather the total
column density \ave{\NHI} from the centre and out, averaged over all
directions, determines the shape of the spectrum.
This leaves us with two unknown parameters, \ave{\NHI} and \Vout, where
\ave{\NHI} is dominated by the number of clouds (\Ncl\,). We first run a rough
fit to the spectrum, providing us with information about the initial conditions for
the system: $\Ncl=10^5$ and $\Vout=150$ \kms, and
consequently run a grid of 11$\times$11 models with $\Ncl\in[10^{4.5},10^{5.5}]$ and
$\Vout\in[100,200]$ \kms .

Instead of doing a regular $\chi^2$ minimization of the pixel-wise difference between
model and spectrum, we compare the observed and the simulated spectra using the following
four observables:
The peak separation in {\AA}; the width of the red peak in {\AA}; the ratio of the integrated
flux in the two peaks; and the ratio of the peak heights. The best fit is defined as the
model which minimizes all of the four above mentioned criteria simultaneously, given the
constraints on metallicity, emission line velocity dispersion, Ly-alpha flux and structure
of the emitting region. From the best
fit, we find an outflow velocity of $\Vout = 160^{+20}_{-40}$~km~s$^{-1}$.
Here, the confidence intervals are given by the range of models, for which {\it all} four
estimators lie within the 68\% confidence interval of those observed in
our spectrum. In the best-fitting model, the average column density and number
of clouds intercepted by a sightline towards the QSO at a distance of 6.3 kpc are
$\log(\NHI / {\rm cm}^{-2})=20.23^{+0.27}_{-1.15}$ and $n_c = 2\pm1$, respectively.
This is somewhat lower although not inconsistent with the measured value
of $20.65\pm0.05$ and the fact that five absorption profiles were used in Sect.~\ref{absfit}.

The escape fraction of \lya\ photons ($f_{\rm esc}\sim90$\%) is higher than what is found when
comparing the total \lya-to-H$\alpha$ ratio, and we find that a SFR of only
6.0 M$_{\odot}$~yr$^{-1}$ is needed to match the observed spectrum.
However, the galaxy was modelled as a spherically symmetric system,
and the galaxy appears disc-like from the {\sc Galfit} analysis in Sec.~\ref{galfit}.
For such a disc-like system, the escape fraction will be significantly lower
when observed edge-on, which indeed seems to be the case here.

\subsection{Star Formation Rate}
\label{sfr}
The F606W flux of the galaxy as measured from the HST data corresponds to a rest frame
wavelength of 1775~{\AA} at the redshift of the galaxy. Based on the AB magnitude obtained
from our {\sc Galfit} analysis (see Table~\ref{GalfitTable}) we get a luminosity after correcting
for Galactic extinction. We use the emission redshift, $z_{em}=2.3537$, to compute the luminosity
distance. This gives a luminosity of
$L_{\nu}=F_{\nu}\, 4\pi\, d_L^2\,(1+z)^{-1} = 9.63\times10^{28}$~erg~s$^{-1}$\,Hz$^{-1}$.
We use the extinction correction factor for the F606W band, $\Delta \mathrm{mag}=-0.11$~mag,
from the NASA/IPAC Extragalactic Database\footnote{http://ned.ipac.caltech.edu/} (NED).
This luminosity corresponds to a
star formation rate of SFR$_{\mathrm{UV}} = 13.5\pm0.5$~M$_{\odot}$~yr$^{-1}$ using the relation
from \citet{Kennicutt98}.

For the H$\alpha$ emission line we take the observed line flux (see table~\ref{FluxTable}), which
corresponds to a luminosity of $L_{\mathrm{H}\alpha} = 2.40\pm0.10\times10^{42}$~erg~s$^{-1}$.
Converting the luminosity into SFR again using \citet{Kennicutt98} gives
SFR$_{\mathrm{H}\alpha} = 18.9\pm0.8~\mathrm{M}_{\odot}$~yr$^{-1}$.

The discrepancy between the two inferred star formation rates indicates some degree of dust extinction.
To quantify the amount of extinction we assume that the two measures should yield the same value,
when corrected for dust reddening. The correction factor to the SFR can be expressed as:
$$ {\rm SFR}_{\mathrm{int}} = {\rm SFR}_{\mathrm{obs}}\cdot 10^{\,0.4\cdot E(B-V) k(\lambda)}\,.$$
\\
Requiring that the two SFRs from UV and H$\alpha$ be equal we arrive at the following expression:
$$ E(B-V) = \frac{2.5}{k(\mathrm{UV})-k(\mathrm{H}\alpha)}
\cdot\log \left( \frac{{\rm SFR}_{\mathrm{H}\alpha}}{{\rm SFR}_{\mathrm{UV}}} \right) ,$$
where $k(\mathrm{UV})$ and $k(\mathrm{H}\alpha)$ denote the \citet{Calzetti2000} extinction
curve evaluated at the UV rest frame wavelength, 1755~\AA, and at the rest frame wavelength
of H$\alpha$, respectively. From this relation we derive a colour excess of $E(B-V) = 0.06\pm0.01$
consistent with the previously mentioned measure from the Balmer decrement (sect.~\ref{balmer}).
The extinction corrected star formation rate is
$\mathrm{SFR}=22.8\pm1.2~\mathrm{M}_{\odot}$~yr$^{-1}$ with an assumed Salpeter initial mass
function (IMF). If we convert this to the Chabrier IMF we get
$\mathrm{SFR_{ch}}=12.7\pm0.7~\mathrm{M}_{\odot}$~yr$^{-1}$. This measurement agrees very well
with the results of \cite{Fynbo10} and \cite{Peroux12} who find SFRs of
SFR $>10$~M$_{\odot}$~yr$^{-1}$ and SFR $=17.1\pm6.0$~M$_{\odot}$~yr$^{-1}$, respectively.

We note that the two proxies for star formation do not trace star formation on the same physical time-scales,
and therefore do not necessarily have to yield the same measured quantity. The H$\alpha$-line primarily
traces the ongoing star formation responsible for ionizing the H{\sc \,ii}-regions, whereas
the UV continuum traces previous star formation as well, linked to the O and B stars already in place.
However, the young, UV-bright stars may also contribute to photoionization of the gas, and the quantification
of the difference depends on assumptions about IMF, star formation history, and the distribution of dust.
The detailed, exact modelling of all these factors is beyond the scope of our simple extinction estimate, we
have therefore chosen to neglect this effect in the above analysis.

The star formation rate can also be determined directly from the absorption lines associated with the DLA
if the C{\sc \,ii}$^*$ absorption line is available \citep[the method is described in][]{Wolfe2003a}. Unfortunately
the line is blended, so we cannot put any firm constraint on the star formation using this method.

\subsection{Broad-Band SED Fitting}
We fit the three broad-band photometric points from the {\it HST} imaging mentioned in Sect.~\ref{galfit} to
obtain estimates of stellar mass, age and star formation rate.
But first, we apply a correction for Galactic extinction available from the NED online database$^2$.
To simplify the fit we subtract the emission-line fluxes from the broad-band fluxes and fit the continuum only,
instead of fitting both continuum and emission lines simultaneously. The only filter which is influenced
by strong emission lines is the F160W band, which contains flux from H$\beta$ and the two
$[$O{\sc \,iii}$]$ lines. In order to subtract the emission lines, we assume a flat continuum spectrum, since
we only detect the emission lines, and subtract the integrated line fluxes from the total flux in the
observed band, weighted by the filter curve.
We find that the emission lines contribute 33\% of the total flux, and we infer a corrected F160W
magnitude of $m_{\mathrm{160}}=23.94\pm0.19$.

Our fitting code uses the stellar population templates from \citet{BC03} convolved with a large
Monte Carlo library of star formation histories (exponential plus random bursts) assuming
a \citet{Chabrier2003} IMF. Dust is added following the two-component model of \citet{Charlot2000},
with the parameters being the total optical depth, $\tau_{\mathrm{v}}$, and the fraction of dust
contributed by the ISM, $\mu$\footnote{For details on the prior distribution of the SFH
and dust parameters see \citet{Salim2005}}. We restrict the range of metallicities of
the models to be consistent with our measurement from the emission lines
$\log(Z/Z_{\odot})=-0.3\pm0.1$. We then adopt a Bayesian approach by comparing
the observed photometry to the one predicted by all the models in the library, and we
construct the probability density functions of stellar mass, mean light-weighted stellar age,
and star formation rate. Taking the mean and 16$^{\mathrm{th}}$ and 84$^{\mathrm{th}}$
percentiles of the PDF we obtain a stellar mass of M$_{\star}=2.1\,^{+1.4}_{-0.9} \times 10^9$~M$_{\odot}$,
age of the galaxy of $t = 98^{+113}_{-48}$~Myr, and a star formation rate of
$\mathrm{SFR}=8.4\,^{+4.3}_{-1.4}~\mathrm{M}_{\odot}$~yr$^{-1}$.
We also get an estimate of the dust extinction from the fit by comparing the intrinsic template with
the best fitting reddened template. From this we infer $A_{\mathrm{V}}=0.08^{+0.29}_{-0.07}$,
which corresponds to $E(B-V)=0.02^{+0.07}_{-0.02}$.

\begin{figure}
  \includegraphics[width=0.48\textwidth]{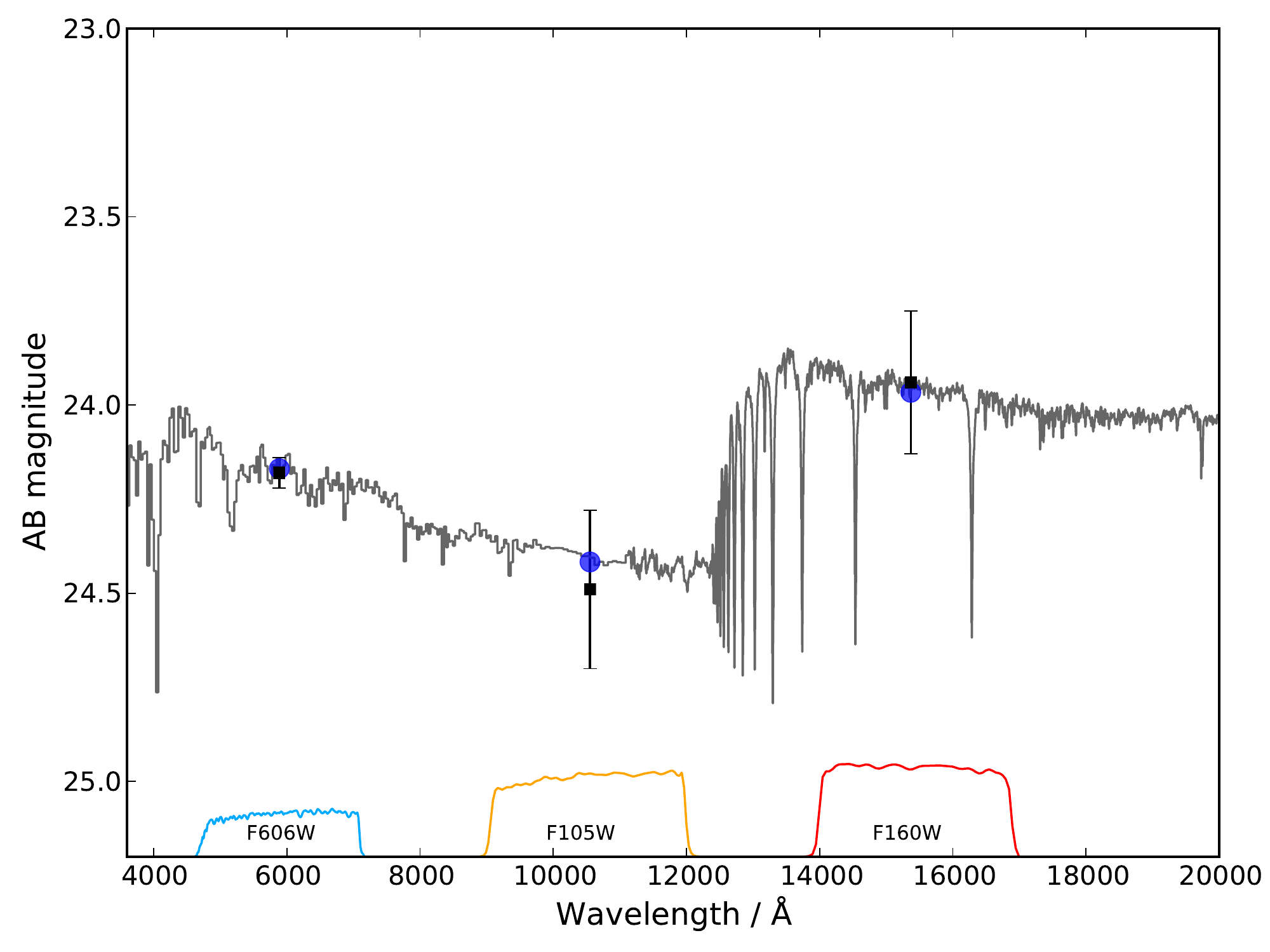}
  \caption{Best fitting template (in grey) to our three broad-band photometric points
  (black, squares). The error-bars in y-direction indicate the 1$\sigma$-uncertainty on the fluxes.
  In the bottom of the figure the transmission curves for each filter are shown, and the blue, round
  points indicate the model magnitudes calculated in each filter passband. The median age and mass
  of the fit is $t = 98$~Myr and M$_{\star}=2.1 \cdot 10^9$~M$_{\odot}$, respectively.
   \label{fig:SED}}
\end{figure}

The SFR from the SED fit agrees well with the one inferred in Sect.~\ref{sfr} within $1 \sigma$, and also
the median value of the dust is consistent with what we find in Sect.~\ref{balmer} and \ref{sfr}.
The photometry and best fitting template are shown in Fig.~\ref{fig:SED}.

\section{Discussion}	

\subsection{Abundances}

We have presented a measurement of the gas-phase metallicity
in the H{\sc \,ii} regions of the emission counterpart of the DLA towards the
quasar Q\,2222$-$0946. The metallicity we infer using the R$_{23}$ and N2 diagnostics
expressed in Solar units is $[\mathrm{O/H}]=-0.30\pm0.13$. 
From the absorbing gas located 6.3~kpc away, we find a metallicity
in the neutral ISM from sulfur of $[\mathrm{S/H}]=-0.49\pm0.05$, thus slightly lower
than the metallicity inferred from the central emitting region. Sulfur is not depleted
onto dust, and thereby traces the overall metallicity very well.
We find a consistent metallicity from $[\mathrm{Si/H}]=-0.54\pm0.05$ and
$[\mathrm{Zn/H}]=-0.38\pm0.05$ (note that ${\rm [Zn/H]}$ is probably overestimated
by 0.1 dex). Neither Zn nor Si are significantly depleted onto dust and hence
provide a good measurement of the gas-phase metallicity \citep{Meyer90, Pettini97}.
The observed metallicities of DLAs at these redshifts range from
$[{\rm M/H}] \approx -2.5$ to $[{\rm M/H}] \approx -0.2$ with an average metallicity
weighted by \NHI\ of $-1.1\pm0.1$ \citep{Rafelski2012}. The DLA in our study is thus amongst the
most metal-rich DLAs at this redshift. 

From the elements Mn, Fe and Ni we clearly see that dust depletion is in fact at play,
the metallicities are $[\mathrm{Fe/H}]=-1.0$, $[\mathrm{Ni/H}]=-0.9$,
and $[\mathrm{Mn/H}]=-1.2$, indicating that the refractory elements are, to some degree,
removed from the gas-phase, which is to be expected in a high-metallicity system
as this particular one \citep{Ledoux03}. Also, we report a tentative detection of
molecular hydrogen, see Sect.~\ref{H2}.

\citet{Bowen2005} find similar results regarding gas phase metallicities
for a low-redshift galaxy, where a quasar intersects the galaxy 3~kpc from
a star forming region. The authors find consistent metallicities
based on the emission region and the absorbing, neutral gas, respectively.
For a compilation of systems with emission and absorption based metallicities,
see \cite{Peroux12}.

The presence of highly enriched material 6~kpc above
(almost perpendicular to) the galactic plane of this galaxy with nearly the same metallicity
as the star forming regions within the galaxy indicates that metal-rich material has been expelled
from the galaxy into the halo. We see independent evidence for outflowing gas from the
Ly$\alpha$ emission line with a velocity of 160~km~s$^{-1}$, see Sect.~\ref{lya-model}
for details. At this velocity we estimate
that it would take of the order of 40 Myr for this enriched material to reach a distance of
6.3kpc from the galaxy plane. Given this relatively short time-scale it seems reasonable
that the two metallicities are similar, since no large amount of enrichment has had time to
occur after the expulsion of the outflowing gas, which is expected to mix with lower
metallicity gas further out, lowering the observed metallicity in the absorbing gas.
This scenario can be compared to the wind observed in the nearby galaxy M82, where
neutral gas and molecules form a filamentary structure in the outflow extending to great
distances from the disc \citep[][]{Veilleux2009, Melioli2013}.

A recent study by \citet{Bregman2013} shows evidence for a similar scenario, where
near-Solar metallicity gas of a nearby, edge-on disc galaxy has been detected 5~kpc above
the disc with a neutral hydrogen column density of $1.3\times10^{20}$~cm$^{-2}$.
However, as these authors study a local galaxy a direct comparison to high redshift
might not be fully valid.

\subsection{Dynamical Mass}
We can use our information about the size of the galaxy and the kinematics, as probed by the emission
lines (see FWHM measures in Table~\ref{FluxTable}), to get an estimate of the dynamical mass of the
system. We follow the method described in \citet{Rhoads2013} to estimate the dynamical mass given
the measured size and velocity dispersion:
$$ M_{\mathrm{dyn}} \approx \frac{4\,\sigma^2\, a_e}{G \sin^2(i)} ~,$$
\\
where $i$ denotes the inclination of the system with $i=90^{\mathrm{o}}$ being edge-on.
In order to estimate the velocity dispersion of the system we use the FWHM of the emission lines as a
probe of the integrated gas-kinematics of the system. We then take
a weighted mean of all the measured line-widths, and correct the FWHM for the instrumental resolution
(45~km~s$^{-1}$). This gives a measure the velocity dispersion of $\sigma=49.1\pm7.7$~km~s$^{-1}$,
and we get the size from the G{\small ALFIT} analysis in physical units:
$a_e=1.12\pm0.06$~kpc (see Table~\ref{GalfitTable}).\\

From the G{\small ALFIT} analysis we infer an axis ratio of the galaxy of $b/a=0.17$. The system may
be described as reasonably disc-like, given the elongated
shape, and the fact that we see a value of S\'ersic $n$ very close to 1.
Using the results of \citet{Haynes1984} that disc galaxies on average have axis ratios around $0.1-0.2$,
we conclude that the galaxy in our study is very close to edge-on, even when assuming the lowest intrinsic
value of $0.1$. We thus adopt a value of $i=90^\mathrm{o}$ and use the fitted half-light semi-major axis
for our estimate of the dynamical mass of the system:
$M_{\mathrm{dyn}} \approx 2.5\times10^{9}$~M$_{\odot}$. This estimate should only be
considered a rough approximation (valid within a factor of $\sim 2$) as we have assumed the system to be
in virial equilibrium, which it may not be.

\subsection{Stellar Mass}
From our SED fit to the broad-band imaging data, we obtain a stellar mass of
M$_{\star}=2.1\times10^9$~M$_{\odot}$. We can use this measurement to test the recently proposed
evolving mass-metallicity relation for DLA systems \citep[][see also \citet{Moller04, Moller2013,
Neeleman2013}]{Ledoux06}. Using the relation in \citet{Moller2013}, which relates emission
metallicity and redshift to stellar mass, with our direct measurement of the emission metallicity,
we find a stellar mass of M$_{\star} = 6\, ^{+9}_{-4} \times 10^9$~M$_{\odot}$. Though the scatter
in their relation is substantial ($\sim0.38$ dex), the agreement between the relation and our best
fit stellar mass from the SED fit is striking.

Moreover, the mass is in very good agreement with the median stellar mass of Lyman Break
Galaxies (LBGs) at redshift $z=1-3$ \citep{Erb06}.
\citet{Hathi2013} find M$_{\star, \mathrm{LBG}}=2\times10^9$~M$_{\odot}$, and
further characterize the general LBG population in terms of median age ($\sim$125~Myr),
median SFR ($\sim$15~M$_{\odot}$~yr$^{-1}$), and median dust extinction
( $E(B-V)\approx$ 0.15). All, but the median dust extinction, agree well
with our inferred quantities. The dust extinction value is, however, still consistent with our value
within the errors \citep[see figure 6 of][]{Hathi2013}.
The consistency between the galaxy in our study and the general LBG population shows that there
is indeed some overlap between galaxies causing DLAs and star-forming galaxies selected with
the Lyman break technique at redshifts $z=2-3$ \citep{Fynbo03, Rauch08}.
Nevertheless, it is important to remember that
the DLA in question was specifically chosen to have high metallicity, and is of unusually high metallicity;
it is an order of magnitude higher than the median metallicity of DLAs at $z=2-3$, $[$M/H$]\simeq -1.5$
or 1/30 of Solar \citep{Noterdaeme08, Rafelski2012}.
The proposed mass-metallicity relation for DLA galaxies \citep{Ledoux06},
then supports the original suggestion by \citet*{Fynbo99} that most DLA galaxy counterparts are
too faint to be identified via their stellar or nebular emission.

We are further able to infer the expected metallicity from the stellar mass and SFR using
the Fundamental Metallicity Relation from the work of \cite{Mannucci10}. Given the
expression from these authors we find an oxygen abundance in the range from
$[{\rm O/H}] = -0.3$ to $[{\rm O/H}] = -0.5$, depending on the fitting function assumed.
Since the SFR of our target is slightly outside the range over which the relation
is derived there will be uncertainty related to the extrapolation.
The metallicity is, however, still in perfect agreement with our measurements. This indicates that
the DLA in our study follows the same relation as other galaxies studied at both lower
and higher redshifts, strengthening the link between this DLA galaxy and the general
population of star-forming galaxies.

\subsubsection*{The Tully-Fisher Relation}
We now turn to look at how this galaxy is located on the stellar-mass Tully-Fisher relation
(M$_{\star}$-TFR) to test our assumption that the system is disc-like and relaxed.
The Tully-Fisher relation, originally stated in terms of luminosity and velocity \citep{Tully1977},
can also be presented in terms of stellar mass (which correlates with luminosity) and velocity.
The M$_{\star}$-TFR was studied in detail by \citet{Kassin2007} who gave their best fit to the data as:
$$ \log(S_{0.5}) = 1.89\pm0.03 +0.34\pm0.05 \cdot \log \left( \frac{M_{\star}}{10^{10}~M_{\odot}} \right)~,$$
\\
where $S_{0.5}$ is defined by the authors as $S_K^2 \equiv K\,V_{\mathrm{rot}}^2+\sigma_g^2$, with $K=0.5$.
Using the measured velocity dispersion as a proxy for $S_{0.5}$ \citep[see a discussion of this in][]{Rhoads2013}
we find that the inferred stellar mass is $M_{\star} = 2.6_{-1.2}^{+1.9} \times 10^{9} M_{\odot}$. This is in
very good agreement with our previously mentioned mass estimates, including our rough estimate of
the dynamical mass.\\

The stellar (and dynamical) mass inferred from the emission line widths is subject to uncertainties caused by
the fact that we do not know the detailed structure of the velocity field. The emission lines are most certainly
influenced by turbulence in the gas, which we cannot quantify. This would overestimate our
line-widths and thereby our TF-based stellar mass and dynamical mass estimates. Also, we required that the
system be in virial equilibrium; however, we observe gas at large galactocentric radii, almost perpendicular
to the disc, with similar metallicity as the line emitting region, indicating outflowing gas from the central parts.
This has an impact on our calculation of the dynamical mass, and our ability to use the line-widths as a
tracer of the ordered rotation of the system. Moreover, a non-negligible gas mass is expected
in a young ($\sim100$~Myr), star-forming galaxy. This mass would
not be accounted for in the stellar mass estimate from the SED fit, but
would contribute to the dynamical mass, thus increasing the observed velocity dispersion.
We estimate the gas mass from the star formation density \citep[see][]{Kennicutt98} using the half-light
radius and the axis-ratio of the galaxy. We infer a gas mass of
M$_{\mathrm{gas}}=1^{+3}_{-0.5}\times10^9$~M$_{\odot}$. This estimate is a very rough approximation
given the large scatter ($\sim0.3$ dex) in the relation, and is therefore not to be trusted as a true value of
the amount of gas. It does, however, indicate that a gas mass of roughly half the stellar mass is present.
We note that the inferred dynamical mass is consistent with our best-fit stellar mass within
$1\,\sigma$, and a significant gas mass is therefore not required in order to reconcile the two mass estimates,
and within the (large) uncertainties, all three mass estimates agree well.
The compact nature of the galaxy also means that the kinematics of the emission lines only probe the
innermost region, and are therefore mostly sensitive to the {\it stellar} mass concentrated in the central
region, and not the gas in the outer parts.

Studies of the TFR at higher redshifts find that the relation is offset to lower stellar masses for a
given velocity. We see a similar though not statistically significant trend in our data.
\citet{Cresci2009} find that LBGs at $z\approx2$ are offset by
0.4 dex compared to the local TFR, and they find that the relation has {\it low} scatter.
\citet{Gnerucci2011} find similar results in their sample of $z\approx3$ galaxies with an offset up to 1 dex,
however, the scatter in their data is very large. This may indicate that the TFR has not yet been established
at these redshifts, and that the galaxies are influenced heavily by random motions.

\section{Summary and Conclusion}
We have presented our analysis of a high-redshift galaxy selected from its neutral hydrogen absorption
seen in the spectrum of a background quasar. We have presented the extracted emission lines from the
galaxy counterpart of the absorption, and combined these with our detailed absorption-line study to probe
the metallicities seen in the two phases. We find that the two metallicities are similar, but the absorbing
gas has a slightly lower metallicity than the emitting gas. We use {\it HST} imaging to constrain the stellar
population; our data are consistent with the picture in which the galaxy causing the Ly$\alpha$ absorption
is a young, small ($\sim1$~kpc), disc-like system, with a not fully ordered (proto-) disc structure. Moreover,
we see evidence for a so-called {\it galactic fountain}, where enriched gas gets blown out from the
star-forming regions, forms the neutral hydrogen absorption that we see $\sim 6$~kpc above the
galactic plane, and in the end may settle back onto the disc.

This galaxy demonstrates exactly how star-forming galaxies at high redshift may overlap with the
population of the most metal-rich DLAs. However, the very faint nature of damped Lyman-$\alpha$
absorbing galaxies renders the majority of these almost impossible to detect. The few exceptions,
such as the case reported here, thus offer a rare glimpse into this elusive galaxy population.

\section*{Acknowledgements}

We thank the anonymous referee for the a nice and constructive
report. The Dark Cosmology Centre is funded by the DNRF. JPUF and PL acknowledge support form
the ERC-StG grant EGGS-278202. JK acknowledges support from an ESO Studentship.
CP has beneÞted from support of the Agence Nationale de la Recherche with reference
ANR-08-BLAN-0316-01. This research has made use of the NASA/IPAC Extragalactic
Database (NED) which is operated by the Jet Propulsion Laboratory, California Institute
of Technology, under contract with the National Aeronautics and Space Administration.
The RT simulations were conducted on the facilities provided by the Danish Center
for Scientific Computing. A.G. acknowledges support from the EU FP7/2007-2013
under grant agreement n. 267251 AstroFIt.

\def\aj{AJ}
\def\araa{ARA\&A}
\def\apj{ApJ}
\def\apjl{ApJ}
\def\apjs{ApJS}
\def\apss{Ap\&SS}
\def\aap{A\&A}
\def\aapr{A\&A~Rev.}
\def\aaps{A\&AS}
\def\mnras{MNRAS}
\def\nat{Nature}
\def\pasp{PASP}
\def\aplett{Astrophys.~Lett.}

\bibliographystyle{mn}

\label{lastpage}

\end{document}